# Omnidirectional ghost imaging system && unwrapping-free panoramic ghost imaging

HUAN CUI,[1] JIE CAO, [1,*] QUN HAO, [1] DONG ZHOU, [1] MINGYUAN TANG, [1] KAIYU ZHANG [2] AND YINGQIANG ZHANG[1]

[1]*School of Optics and Photonics, Beijing Institute of Technology, Beijing 100081, China*
[2]*School of Sports Engineering, Beijing Sport University, Beijing 100084, China.*
*\*ajieanyn@163.com*

**Abstract:** Ghost imaging (GI) is a novel imaging method, which can reconstruct the object information by the light intensity correlation measurements. However, at present, the field of view (FOV) is limited to the illuminating range of the light patterns. To enlarge FOV of GI efficiently, here we proposed the omnidirectional ghost imaging system (OGIS), which can achieve a 360° omnidirectional FOV at one shot only by adding a curved mirror. Moreover, by designing the retina-like annular patterns with log-polar patterns, OGIS can obtain unwrapping-free undistorted panoramic images with uniform resolution, which opens up a new way for the application of GI.

## 1. Introduction

Unlike the traditional imaging via a pixelated detector array, ghost imaging (GI) is a novel imaging method using a bucket detector with no spatial resolution, which can reconstruct the object information by means of the light intensity correlation measurements whether with an entangled photon source or a classical source [1-5]. Due to its nonlocal imaging property and high detection efficiency, GI can solve many problems from traditional imaging techniques, which has been widely used in the fields of remote sensing [6], multispectral imaging [7], microscopy [8], optical encryption [9] and so on. However, at present, the field of view (FOV) of GI is limited to the illuminating range of the light source or digital mirror device (DMD), which cannot meet the imaging requirements of large FOV or even 360° view in many vision fields, such as security monitoring [10], visual navigation [11], medical diagnosis [12]. Besides, with respect of practical use in special spectral application, e.g., infrared reconnaissance, large FOV requires pixelated array, which results in high cost and difficult manufacture. Therefore, in order to expand the practical application fields of GI, it is necessary to extend the illuminating range of the light source or DMD and further extend the FOV of GI. In this paper, we introduce the catadioptric omnidirectional imaging method [13] into the GI system, called as omnidirectional ghost imaging system (OGIS), where the scene can be illuminated with a 360° omnidirectional FOV at one shot only by adding a curved mirror with a strong reflective property.

Due to the design freedom of light patterns of GI displayed on DMD [14-17], OGIS can obtain unwrapping-free undistorted panoramic images with uniform resolution, which solves many of challenges in the traditional catadioptric omnidirectional imaging system. For one thing, since the geometry of a curved mirror is usually convex, omnidirectional images are usually circular and distorted. Besides, the images sampling by traditional imaging device like CCD with uniform resolution always leads to a non-uniform resolution, where the projected resolution quality decreases on a gradient from the periphery to the center [18]. Therefore, inspired by the retina of human eye [19], a retina-like annular pattern is designed. The log-polar structure of the designed annular pattern can not only match the convex geometry of a curved mirror, but also solve the non-uniform sampling resolution from the center to the periphery to further obtain omnidirectional images with uniform circumferential angular resolution. For another, the unwrapping process seems to be necessary to obtain a rectangular panoramic image, because omnidirectional images with concentric annular distortion are not adapted to the human visual system. However, similar to the unwrapped panoramic image obtained by the log-polar transform [20], a novel unwrapping-free panoramic GI is proposed. Due to the log-polar structural characteristic of the retina-like annular pattern, undistorted panoramic ghost images can be reconstructed directly by using the corresponding rectangular patterns mapped in log-polar coordinate, which not use the unwrapping process. The results are beneficial to improving the performances of GI in larger FOV, which paves the way for practical use.

## 2. Methods

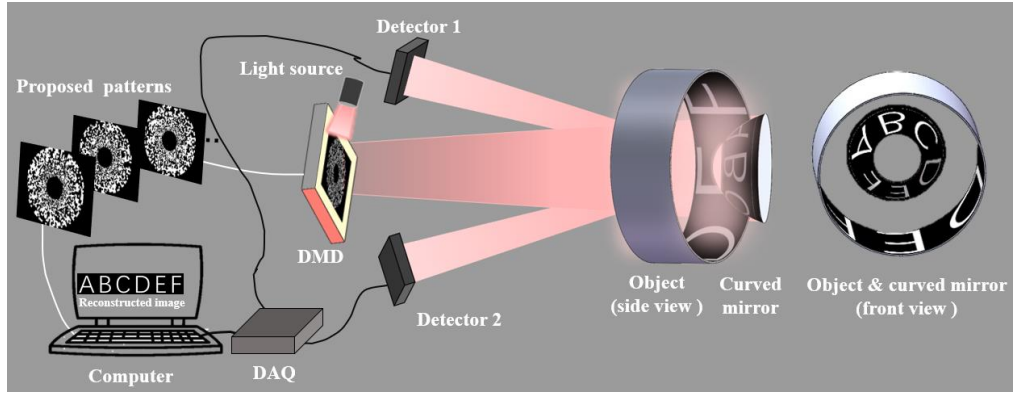

Fig. 1. Schematic of OGIS.

Compared with the traditional GI system with a structured illumination configuration [25], the proposed OGIS can achieve a 360° omnidirectional FOV with a simple structure by only adding a curved mirror at the object location, as shown in Fig. 1. The DMD projects the light pattern onto the curved mirror, due to the multi-reflection of light between the curved mirror and the object surface, the illuminating range of the light is expanded extremely, and then all of the scenes with 360° omnidirectional FOV are illuminated. However, there is a difference between OGIS and the traditional omnidirectional imaging system, where all the back-scattered light intensity from all of the scenes with 360° omnidirectional FOV should be collected, rather than just the one from the curved mirror. Therefore, in order to collect more light intensity uniformly from all the direction, we employ two bucket detectors in the present imaging system for the limited detection range of one single-pixel detector.

Moreover, in order to obtain the scene with 360° omnidirectional FOV conveniently, the object is made by a hollow cylinder attached to a string of letters 'ABCDEF'. Form the front view of object & curved mirror in Fig. 1, we can clearly see that the image on the curved mirror is circular and distorted. Therefore, in order to match the imaging features of the curved mirror, we design a retina-like annular pattern based on our previous proposed retina-like log-polar structure [15], which includes $P$ rings with increasing radius outside the fovea blind hole, and each ring is divided into $Q$ cells by angle uniformly, as shown in Fig. 2(b). And the mathematical model is expressed as:

$$\begin{cases} r_p = r_0 \cdot \varepsilon^p \\ rc_p = rc_1 \cdot \varepsilon^{p-1} \\ \varepsilon = \dfrac{1+\sin(\pi/Q)}{1-\sin(\pi/Q)} \\ rc_1 = \dfrac{r_0}{1-\sin(\pi/Q)} \\ \theta_q = q \cdot \dfrac{2\pi}{Q} \qquad (q=1,2,3...Q) \\ \xi_p = \log_\varepsilon(rc_p) = \log_\varepsilon(rc_1) + p-1 \qquad (p=1,2,3...P) \end{cases} \qquad (1)$$

where $r_p$ is the outer radius of the $p$th ring, $rc_p$ is the center radius of the $p$th ring, $r_0$ is the radius of the fovea blind hole, $\varepsilon$ is the increasing coefficient of the radius between adjacent rings, $\theta_q$ is the angle of the $q$th cell of each ring, $\xi_p$ is the value of the center radius of the $p$th ring in log-polar coordinates.

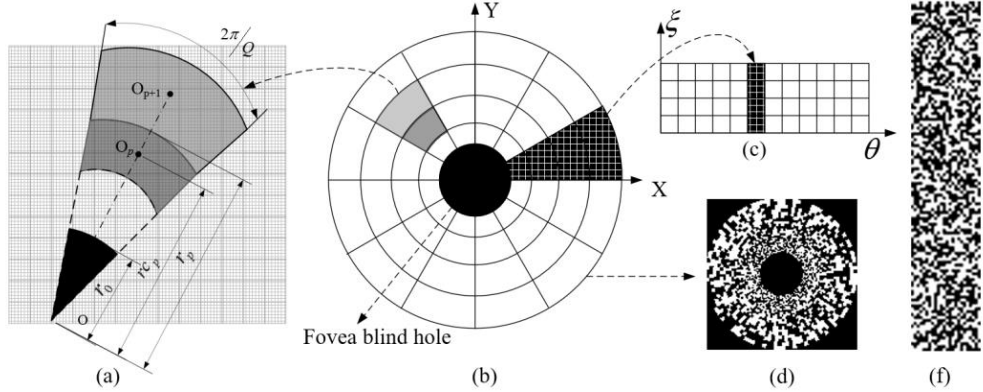

Fig. 2. Retina-like annular pattern with log-polar structure used in OGIS. (a) The detail of the retina-like annular pattern by matching the log-polar structure with a grid with uniform resolution fittingly; (b) The log-polar structure of the retina-like annular pattern in the Cartesian coordinate system, and the central area with the radius $r_0$ is the fovea blind hole, with the corresponding micromirrors states 'Off' on the DMD; (c) The detail of the uniform rectangular structure of the annular pattern mapped in Log-polar coordinate system; (d) An example random retina-like annular pattern with 24 rings divided into 120 cells, displayed on DMD with 128×128 pixels; (f)The uniform rectangular pattern with 24×120 pixels mapped in Log-polar coordinate system of the retina-like annular pattern shown in (d).

In our OGIS, in order to display fittingly the retina-like annular pattern on the DMD made up with a uniform array of micromirrors, as shown in Fig. 2(a)&(d), the retina-like annular pattern is created by matching the log-polar structure with a uniform-resolution square grid, which can be viewed as the circumscribed square of the outermost ring with the radius $r_P$. Therefore, a retina-like annular pattern displayed on the DMD can be denoted as $P_{annular}$ with $m \times m$ pixels, and the light intensity $I$ after projecting the pattern $P_{annular}$ onto the curved mirror can be given by:

$$I = \sum_{x=1}^{m}\sum_{y=1}^{m} P_{annular}(x,y) \cdot O_{omnidirectional}(x,y), \qquad (2)$$

where $x$ and $y$ index the position of the retina-like annular pattern on the square grid in Cartesian coordinate system, and $O_{omnidirectional}$ is the omnidirectional object reflected onto the curved mirror.

In order to reconstruct the omnidirectional image by less measurements and higher quality, the compress sensing (CS) reconstruction algorithm based on the total variation (TV) [21] is used in our OGIS, which transforms the image reconstruction problem into a constrained optimization problem. Mathematically, we use $l_1$ norm to calculate the total variation of the omnidirectional object, with the gradient transformation matrix $H$ and the corresponding coefficient vector $c$, the optimization model is as follows:

$$\begin{aligned} \min \quad & \|c\|_{l_1} \\ s.t. \quad & HO_{omnidirectional} = c \\ & P_{annular} O_{omnidirectional} = I \end{aligned} \qquad (3)$$

where $\boldsymbol{P}_{annular} \in R^{T \times M}$ is the annular pattern matrix (the number of projected patterns is $T$, each pattern consists of $M = m \times m$ pixels), $\boldsymbol{O}_{omnidirectional} \in R^{M \times 1}$ is the omnidirectional object discretized by $M$ pixels, and $\boldsymbol{I} \in R^{T \times 1}$ is the measurement matrix corresponding to $T$ projected patterns. And finally, the omnidirectional ghost image $\boldsymbol{O'}_{omnidirectional}$ with $m \times m$ pixels is reconstructed.

Further, interestingly, as shown in Fig. 2(f), due to the log-polar structural characteristic of the retina-like annular pattern, the corresponding pattern mapped in Log-polar coordinate system is rectangular and uniform-resolution with $P \times Q$ pixels, which is similar to a rectangular panoramic image unwrapped from an omnidirectional image by the log-polar transform [20]. Therefore, inspired by the similarity, we can reconstruct the panoramic image directly by rectangular patterns mapped in log-polar coordinate without the unwrapping process. Detailly, the optimization model of TV is change by:

$$\begin{aligned} \min \quad & \|c\|_{l_1} \\ s.t. \quad & HO_{panoramic} = c \\ & P_{rectangular} O_{panoramic} = I \end{aligned} \quad (4)$$

where $P_{panoramic} \in R^{T \times N}$ is the corresponding rectangular pattern matrix mapped in Log-polar coordinate system (each pattern consists of $N = P \times Q$ pixels), $O_{panoramic} \in R^{N \times 1}$ is the wished panoramic object, and it is noted that the measurement matrix $I$ is the same as the one used for reconstructing the omnidirectional ghost image. And finally, the unwrapping-free panoramic ghost image $O'_{panoramic}$ with $P \times Q$ pixels is reconstructed directly.

## 3. Results

### 3.1 Experimental setup

According to Fig. 1, the experimental setup of OGIS includes the projection part, the detection part, the object and a curved mirror. The projection part includes a light emitting diode (LED) operating at 500-700nm (@20W), DMD (Texas Instruments DLP4100, 1024×768 micromirror array) and a projection lens with $f = 150$mm. The detection part consists of two same photodetectors (Thorlabs PDA36A, 13 mm$^2$ active area) and the data acquisition board (DAQ) (PICO6404E, sampling at 1 MS/s). The object with 360 omnidirectional FOV is made by a hollow cylinder (the height is 50mm and the base diameter is 75mm) attached to a string of letters printed on A4 paper. And a plane-convex lens coated with a strong reflective film (the diameter is 50.8mm, the central thickness is 16.3mm and the radius of curvature is 30.91mm) is used as the curved mirror for the experiment.

Based on the mentioned experimental setup of OGIS, it is necessary to choose suitable parameters of the retina-like patterns to achieve the omnidirectional GI and panoramic GI, such as $r_0$, $P$, $Q$. Firstly, it should be noted that there is a self-reflection of the projection part on the center of the curved mirror, which is not the wished object, and therefore, the fovea blind hole in the retina-like annular pattern is used to cover the related area. According to the geometric image relationship between the curved mirror and the projection part [22], the self-reflected area of the projection part on the curved mirror can be calculated, and correspondingly, the ratio of the fovea blind hole to the total annular pattern on the OGIS is obtained $\eta = 28.9\%$, which can be expressed as $\eta = r_0/r_P$. In the experiments of the paper, the retina-like annular pattern displayed on the DMD is designed with 128×128 pixels, that is, take a pixel as a unit resolution of 1, then $r_P = m/2 = 64$, $r_0 = \eta \times r_P \approx 18.5$. In addition, from the equation (1), we can get the relationship between $P$ and $Q$, given by:

$$P = \log_\varepsilon^\eta + \log_\varepsilon^{(1-\sin\frac{\pi}{Q})} \quad \text{where } \varepsilon > 1, Q > 2, \text{and } P, Q \in \mathbb{N} \quad (5)$$

where $\eta$ is a constant in our OGIS. Mathematically, $P$ is larger with the increasing of $Q$. In order to improve the imaging quality, it is necessary to obtain more rings divided into more cells, which represents the sampling resolution of the retina-like annular patterns. However, because of the limitation of the minimum resolution, the imaging quality is not always better with the increasing rings and cells. In order to avoid the over resolution that there will be several cells corresponding to one pixel, we choose the largest integral one with the condition that $r_1-r_0>1$, that is, P=24, Q=120. Therefore, the retina-like annular pattern can be designed detailly with the mentioned parameters setting, and an example is shown is Fig. 2(d).

### 3.2 Omnidirectional GI

With the proposed OGIS, omnidirectional ghost images can be obtained, as shown in Fig. 3. Firstly, to verify the optimal choice of parameters $P$ and $Q$, according to the equation (5), we project the retina-like annular patterns with different $P$ and $Q$ to reconstruct the omnidirectional ghost images of the letters 'ABCDEF', as shown in Fig. 3(a). It is clearly seen that, at the beginning, with the increasing of $P$ from 5 to 24, the sampling resolution is higher, and the imaging quality is better with the clearer image. However, with the existing over resolution that P>24, the imaging quality deteriorates slightly. Expeically, as the over resolution exists all of the rings, like the one with P=100 and Q=500, the corresponding retina-like annular pattern can be viewed as a annular random pattern with the mimmun resolution. Further, compared to the reference image obtained by the traditional camera, the calculated PSNR also shows that the omnidirectional ghost image with P=24 and Q=120 is the one with the best imaging quality. And then, using the same retina-like annular patterns with P=24 and Q=120, we imaged the object with the Arabic numerals and the Chinese characters similarly, as shown in Fig. 3(b).

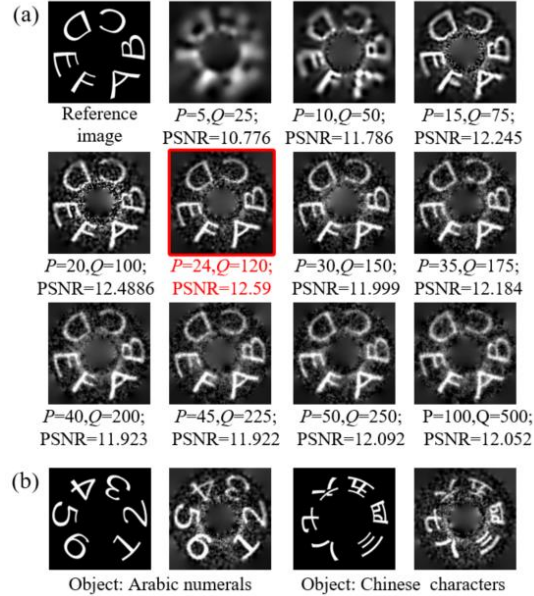

Fig. 3. Omnidirectional ghost images obtained by OGIS. (a) The omnidirectional ghost images of the letters 'ABCDEF' reconstructed by retina-like annular patterns with different $P$ and $Q$ respectively. (b) The omnidirectional ghost images of the Arabic numerals '123456' and the Chinese characters '三四五六七八' reconstructed by retina-like patterns with $P=24$ and $Q=120$ respectively, where the left one is the reference image, and the right one is the reconstructed image.

### 3.3 Unwrapping-free panoramic GI

Since the omnidirectional ghost images with concentric annular distortion are not adapted to the human visual system, panoramic ghost images are required conveniently for the next step of image processing in many vision fields. Generally, with the traditional unwrapping process, the panoramic ghost images can be obtained by the log-polar transformation from the reconstructed omnidirectional ghost images, which corresponds to the log-polar structure of the projected retina-like annular patterns, as shown in Fig. 4(a). Moreover, here is another method proposed to obtain the panoramic ghost images without the unwrapping process. Inspired by the log-polar transformation from the omnidirectional images to the panoramic images, a novel unwrapping-free panoramic GI is proposed by the direct reconstruction of GI using rectangular patterns with P×Q pixels, which are the mappings of the corresponding retina-like annular patterns in the Log-polar coordinate system, as shown in Fig. 4(b). In order to observe the imaging effects of the proposed two methods, panoramic ghost images are obtained respectively with different measurements T=1650 and T=2880, as shown in Fig. 4(c)-(f). Subjectively, the imaging effects of the two methods are similar to each other. However, in comparison with the unwrapping-free panoramic images, the one using the traditional unwrapping process from the omnidirectional image is noisier, which is also clear seen from PSNR and SSIM.

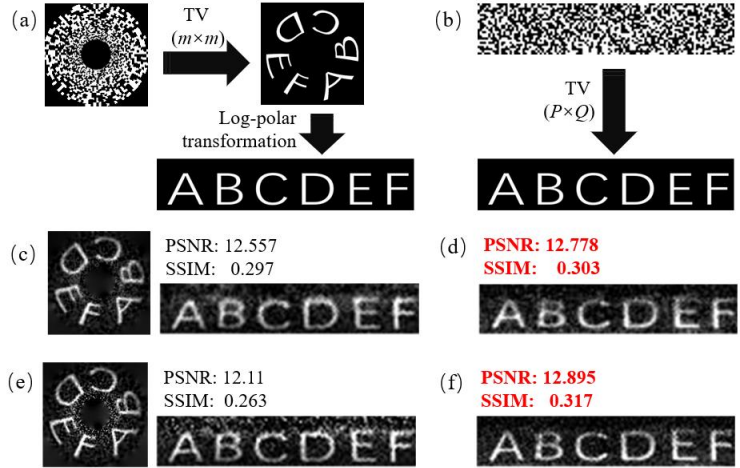

Fig. 4. Unwrapping-free panoramic ghost images & unwrapping panoramic ghost images. (a) The schematic diagram of the unwrapping panoramic GI; (b) The schematic diagram of the unwrapping-free panoramic GI; (c) and (e) are the panoramic ghost images with the method shown in (a) with different measurements T=1650 and T=2880 respectively; (d) and (f) are the corresponding panoramic ghost images with the method shown in (b) respectively.

In addition, during the reconstruction process by using TV algorithm, the resolution of the rectangular patterns to obtain the panoramic image is $P×Q$, while the resolution of the corresponding retina-like patterns to obtain the omnidirectional image is $m×m$. Since $m×m$ is much larger than $P×Q$, take an example of 128×128 > 24×120, the calculated amount and time of the direct reconstruction with rectangular pattens is reduced extremely, which can provide a valuable approach to the real-time panoramic ghost image.

### 4. Conclusions

In order to expand the practical application fields of GI, whose imaging requirements are large FOV or even 360° view, inspired by the catadioptric omnidirectional imaging method, OGIS is proposed with 360° omnidirectional FOV only by adding a curved mirror with a strong reflective property. Firstly, the retina-like annular pattern with log-polar structure is designed to match the geometry of the curved mirror as well as obtain the uniform sampling resolution over the whole FOV. In addition, the related parameters are chosen to obtain the omnidirectional ghost images with the best imaging quality. Further, to obtain the panoramic ghost images adapted to the human eye vision system, we gave the two methods of panoramic GI. One is the unwrapping panoramic GI based on the traditional unwrapping process, where the panoramic images are obtained by the log-polar transformation from the reconstructed omnidirectional ghost images. Another is a novel unwrapping-free panoramic GI, where the panoramic images can be reconstructed directly using rectangular patterns and the same measurement sequence as the one to reconstruct the omnidirectional ghost images. It is noted that the rectangular patterns are the mappings of the corresponding retina-like annular patterns in Log-polar coordinate system. From the experimental results, the proposed unwrapping-free panoramic GI has a little better imaging quality than the traditional one with the unwrapping process, and the processing time and data amount is reduced extremely, which can provide a meaningful approach to achieve real-time panoramic ghost imaging.